%% file: main.tex
\RequirePackage{fix-cm}
\documentclass[smallextended]{svjour3}       

\usepackage{booktabs} 

\usepackage{paralist}
\usepackage{etoolbox}

\include{includes}

\include{macros}
\include{pygments}

\usepackage{epstopdf}
\lstdefinestyle{example} {
	frame=tb,
	basicstyle=\footnotesize\ttfamily,
	numbers=left,
	language=C
}

\begin{document}

\title{A Case Study on the Impact of Similarity Measure on Information Retrieval based Software Engineering Tasks}

\author{Md Masudur Rahman \and
        Saikat Chakraborty \and
        Gail Kaiser \and
        Baishakhi Ray
}


\institute{Md Masudur Rahman \at
              University of Virginia \\
              \email{masud@virginia.edu}           
         \and
         Saikat Chakraborty \at
              University of Virginia \\
              \email{saikatc@virginia.edu} 
          \and
         Gail Kaiser\at
              Columbia University \\
              \email{kaiser@cs.columbia.edu} 
          \and
           Baishakhi Ray \at
              Columbia University \\
              \email{rayb@cs.columbia.edu} 
}

\date{Received: date / Accepted: date}

\maketitle

\begin{abstract}
Information Retrieval (IR) plays a pivotal role in diverse Software Engineering (SE) tasks, \eg bug localization and triaging, code retrieval, requirements analysis, \etc The choice of similarity measure is the core component of an IR technique. The performance of any IR method critically depends on selecting an appropriate similarity measure for the given application domain. Since different SE tasks operate on different document types like bug reports, software descriptions, source code, \etc that often contain non-standard domain-specific vocabulary, it is essential to understand which similarity measures work best for different SE documents.

This paper presents two case studies on the effect of different similarity measure on various SE documents \wrt two tasks: (i) \textit{\projecttask}: finding similar \gh projects and (ii) \textit{\bugtask}: retrieving buggy source file(s) correspond to a bug report. These tasks contain a diverse combination of textual (\ie description, readme) and code (\ie source code, API, import package) artifacts.  
We observe that the performance of IR models varies when applied to different artifact types. We find that, in general, the context-aware models achieve better performance on textual artifacts. In contrast,  simple keyword-based bag-of-words models perform better on code artifacts. On the other hand, the probabilistic ranking model BM25 performs better on a mixture of text and code artifacts.

We further investigate how such an informed choice of similarity measure impacts the performance of SE tools. In particular, we analyze two previously proposed tools for \projecttask and \bugtask tasks, which leverage diverse software artifacts, and observe that an informed choice of similarity measure indeed leads to improved performance of the existing SE tools.   

\end{abstract}


\input{paper.tex}


\bibliographystyle{spphys}       
\bibliography{main}   

\end{document}

%% file: includes.tex
\usepackage{fancybox}
\usepackage{graphicx}
\graphicspath{{figures/}}
\DeclareGraphicsExtensions{.pdf,.jpeg,.png}
\usepackage{amsmath}
\usepackage{caption}
\usepackage{fancyvrb}
\usepackage{color}
\usepackage{algorithmic}
\usepackage{array}
\usepackage{mdwmath}
\usepackage{mdwtab}
\usepackage{booktabs}
\usepackage{listings}
\usepackage{eqparbox}
\usepackage{stfloats}
\usepackage{url}
\usepackage{comment}
\usepackage{framed}
\usepackage{ifthen}
\usepackage{balance}
\usepackage{enumitem}

\usepackage[turnoff]{notes}

\usepackage{multirow}
\usepackage{xspace}
\usepackage{listings}
\usepackage{subfig}

\newcommand{\rom}[1]{\uppercase\expandafter{\romannumeral #1\relax}}

\newcommand{\etal}{\hbox{{et al.}}\xspace}
\newcommand{\eg}{\hbox{{e.g.,}}\xspace}
\newcommand{\ie}{\hbox{{i.e.}}\xspace}

\newcommand{\wrt}{\hbox{{w.r.t.}}\xspace}

\newcommand{\etc}{\hbox{{etc.}}\xspace}

\newcommand{\tableshrink}[1]{\vspace{-0.25in}}

\definecolor{gray50}{gray}{.5}
\definecolor{gray40}{gray}{.6}
\definecolor{gray30}{gray}{.7}
\definecolor{gray20}{gray}{.8}
\definecolor{gray10}{gray}{.9}
\definecolor{gray05}{gray}{.95}

\newlength\Linewidth
\def\findlength{\setlength\Linewidth\linewidth
\addtolength\Linewidth{-4\fboxrule}
\addtolength\Linewidth{-3\fboxsep}
}
\newenvironment{examplebox}{\par\begingroup
   \setlength{\fboxsep}{5pt}\findlength
   \setbox0=\vbox\bgroup\noindent
   \hsize=0.95\linewidth
   \begin{minipage}{0.95\linewidth}\normalsize}
    {\end{minipage}\egroup
    \textcolor{gray20}{\fboxsep1.5pt\fbox
     {\fboxsep5pt\colorbox{gray05}{\normalcolor\box0}}}
    \endgroup\par\noindent
    \normalcolor\ignorespacesafterend}

\newcounter{RQCounter}
\newcounter{RQACounter}

\newcommand{\RQA}[2]{%
\refstepcounter{RQACounter} \label{#1}
\vspace{0.1in} \noindent\textbf{RQ\arabic{RQACounter}.~#2 \vspace{0.05in}}

}

\usepackage{tcolorbox}
\definecolor{mycolor}{rgb}{0.122, 0.435, 0.698}
\newcommand{\mybox}[1]{%
  \setbox0=\hbox{#1}%
  \setlength{\@tempdima}{\dimexpr\wd0+13pt}%
  \begin{tcolorbox}[colframe=mycolor,boxrule=0.5pt,arc=4pt,
      left=6pt,right=6pt,top=6pt,bottom=6pt,boxsep=0pt,width=\@tempdima]
    #1
  \end{tcolorbox}
}

\newcommand{\RS}[2]{%
\begin{tcolorbox}[colback=white,boxrule=0.5pt,arc=4pt]
\textbf{\underline{Result {\ref{#1}}}:~}{{\em #2}}%
\end{tcolorbox}
}

\definecolor{javared}{rgb}{0.6,0,0} 
\definecolor{javagreen}{rgb}{0.25,0.5,0.35} 
\definecolor{javapurple}{rgb}{0.5,0,0.35} 
\definecolor{javadocblue}{rgb}{0.25,0.35,0.75} 

\lstdefinestyle{customc}{
  belowcaptionskip=\baselineskip,
  breaklines=true,
  xleftmargin=\parindent,
  language=java,
  showstringspaces=false,
  basicstyle=\scriptsize\ttfamily,
  keywordstyle=\bfseries\color{javapurple},
  commentstyle=\itshape\blue,
  belowskip=-10pt,
  aboveskip=-5pt
}

%% file: macros.tex
\newcommand{\projecttask}{project recommendation\xspace}
\newcommand{\ProjectTask}{Project Recommendation\xspace}
\newcommand{\bugtask}{bug localization\xspace}
\newcommand{\BugTask}{Bug Localization\xspace}

\newcommand{\simprojdata}{1590\xspace} 
\newcommand{\simprodatacategory}{78\xspace}

\newcommand{\gh}{GitHub\xspace}

\newcommand{\md}[1]{{\scriptsize \todo{Masud:  {\color{blue} #1}}}}
\newcommand{\bray}[1]{{\scriptsize \todo{Bray:  {\color{red} #1}}}}

\newcommand{\DES}{description\xspace}
\newcommand{\RME}{readme\xspace}


%% file: pygments.tex
\newcommand\at{@}

\newcommand\blue[1]{\textcolor[rgb]{0.00,0.00,1.00}{{#1}}}

%% file: paper.tex

\section{Introduction}
\label{sec:intro}	

Information retrieval (IR) techniques play a pivotal role in many Software Engineering (SE) tasks. Haiduc \etal identified more than 20 different SE tasks~\cite{Haiduc2016}, \eg feature location, traceability link recovery,  bug localization and triaging \etc, that use IR or Text Retrieval (TR) techniques\footnote{TR  is a branch of IR~\cite{arnaoudova2015use}. We use the terms interchangeably.}. 
Typically, an IR method depends on two key components: (i) {\em query}: contains users' information needs, and (ii) {\em candidate document}: from which the relevant information is extracted. Given a query, an IR task finds relevant candidate documents by computing a similarity score between the query and each document using an appropriate similarity measure. 
The candidate documents are then ranked based on the decreasing values of the similarity scores. Thus, measuring document similarity is one of the key components of any IR technique. Exact matching (\eg Boolean), matching based on similar bag-of-words (\eg VSM~\cite{salton1975vector}), context-based matching (\eg LSI~\cite{landauer1997solution}, WMD~\cite{kusner2015word}), \etc are a few popular methods for measuring document similarities. 

Usually, well-established IR techniques are used to measure document similarities for SE tasks, primarily because they are already stable, fine-tuned, and well-explored. However, these models are refined mainly for natural language (NL) text corpora. SE corpora, which usually contain a diverse set of information including source code, bug reports, project descriptions, API documentations, \etc, are linguistically different from natural language~\cite{hellendoorn2017deep} due to the following properties: 
(i) {\em less ambiguous}: Documents containing source code are {less ambiguous} than NL so that compiler can interpret the code. 
(ii) {\em open vocabulary}: Source code contains {open vocabulary} where developers can coin new variable names without changing the semantics of the programs. 
(iii) {\em diverse}: Different types of SE documents can have different language properties. For example, a bug report that primarily contains NL text with domain-specific keywords is linguistically very different from source code or execution traces. Thus, while applying the IR models to SE context, we have to adapt the models to cater these SE characteristics\textemdash a best performing similarity measure, which is tuned for NL tasks, may not be optimal for SE corpora. In fact, different similarity measures might be suitable for different type of SE artifacts. 

In this paper, we extensively evaluate how the performances of SE tasks vary with the choices of different similarity measures. In particular, we present two case studies: 
(i) {\em \projecttask}: given a \gh project as query, it tries to find functionally similar \gh projects, and 
(ii) {\em \bugtask}: given a bug-report as query, it retrieves relevant buggy source file(s). We carefully choose these tasks as representative SE jobs because they involve contrasting diverse sets of SE artifacts. In particular, for \projecttask, the similarity is measured among different textual  (\eg text vs.~text) and code (\eg code vs.~code) artifacts separately, whereas the latter relies on the similarity computation between mixtures of textual and code document types (\eg code vs.~text). 

We evaluate the effectiveness of four popular IR measures: Vector Space Model (VSM)~\cite{salton1975vector}, Latent Semantic Indexing (LSI)~\cite{landauer1997solution}, BM25~\cite{robertson1994some}, and embedding based Word Mover's Distance (WMD)~\cite{kusner2015word}  on different SE documents.  Each of these measure has its own benefit and therefore has its preferred domain.  For example,  in SE literature, VSM~\cite{ye2014learning,ye2016mapping,wen2016locus,zhou2012should,niu2017learning} and LSI ~\cite{mcmillan2012detecting,mcmillan2010recommending,marcus2004information} are commonly used,  while BM25 is popular in general purpose search engines, and WMD is the state-of-the-art similarity measure for natural language document classification. Recently, researchers start proposing word embedding based models to improve different SE tasks  ~\cite{ye2016word,van2016characterizing,xu2016predicting}. Thus, we choose different measures that are known to be effective for different corpus and tasks, and analyze them thoroughly for SE artifacts.

We study $1832$ \gh projects for the \projecttask task and $1100$ bug reports for the \bugtask task. 
We observe that context-aware models such as LSI and WMD are in general better, while the keyword based bag-of-words (BOW) model VSM performs best for code-only artifacts. In contrast, BM25 performs the best on mixed artifacts. Surprisingly, BM25 is not as effective for text- or code-only documents. 

To further investigate the impact of such an informed model choice, we conduct a study on two previously proposed tools: CLAN~\cite{mcmillan2012detecting} for \projecttask and LR~\cite{ye2014learning} for the \bugtask task. CLAN leverages LSI to compute the similarity between two type of code artifacts: package and API. We replaced the LSI similarity measure with VSM, as we observe that the VSM model is optimal (among the IR models considered) for code-only artifacts. Experiments on our dataset confirmed that the informed model choice indeed improves the performance of this \projecttask tool.
We conducted a similar experiment on the \bugtask task. Ye \etal's bug localization tool LR~\cite{ye2014learning} leverages VSM to compute the similarity between a bug report and source code (\ie mixture of text and code). Since our experiments show that BM25 is the optimal model choice (among IR models considered) for such mixtures, we then replace the VSM similarity measure with BM25. We observe that our modification indeed improves the performance of the \bugtask on a benchmark dataset. 

We make following contributions:
\begin{enumerate}[leftmargin=*]
\item 
We evaluate four IR-based similarity measures on diverse artifacts \wrt two SE tasks.
\item
 We provide empirical evidence that an optimal choice of similarity measure can improve the accuracy of these two IR-based SE tasks. 

\item
We curate a valuable dataset of $1832$ \gh projects by retrieving their descriptions, readme contents, method class names, imported package usage, and API for \projecttask. We manually associate each project with a fine-grained category that describes their functionalities. Our dataset is available open source at \url{https://drive.google.com/drive/folders/1wqViscmoD_ikFuxxbcF0aTHD3gvnWWw4?usp=sharing}.

\end{enumerate}

 The rest of the paper is organized as follows: Section~\ref{sec:back} presents relevant technical background. Next, in Section~\ref{sec:method}, we discuss methodology for \projecttask and \bugtask tasks, including data collection, feature extraction, and evaluation metric. We discuss our case study in Section~\ref{sec:case-study} \wrt the two SE tasks including research questions, experimental results, and implications.
In Section~\ref{sec:discussion}, we summarize our experimental findings with possible implications and future works. We discuss related works in Section ~\ref{sec:related}. Finally, in Section~\ref{sec:threats}, we examine potential threats to validity that may affect our findings, and we conclude in Section~\ref{sec:conc}.

\section{Background}
\label{sec:back}
In this section, we present the relevant technical background.

\subsection{Similarity Measure}

\textbf{1.~Vector Space Model (VSM).}
In VSM, documents (D) and queries (q) are represented as N-dimensional vectors where N is the size of the vocabulary, and each dimension corresponds to a separate word or term. Each vector element represents the weight of the corresponding term; \ie, $q = (qw_1,qw_2,qw_3,...,qw_N)$ and $D = (Dw_1,Dw_2,Dw_3,...,Dw_N)$ where the $qw_i$ and $Dw_i$ are the weights of the term $i$ in a "bag-of-word" representation of vocabulary size $N$.
An effective way to compute the term weight is the {term frequency-inverse document frequency} (tf-idf), where tf represents the importance of the term in a document, and idf represents how valuable or rare the term is across all the documents. 
Then the similarity between two documents is computed as the cosine angle between corresponding vectors as $sim(q, D) = \cos(q, D) = \frac{q^TD}{||q||\ ||D||}$. The high cosine value means the two documents are similar.

{\em Implication:} VSM model is effective and simple to implement. However, since it is a BOW and exact keyword matching approach which ignores the order of the tokens in a document, this method is suitable when the order of the words does not matter. 
Note that in the cosine similarity formula, the magnitudes of the document vectors ($||q||$ and $||D||$) are in the denominator and give smaller cosine value for the larger dimensional vector. Thus, longer documents may be penalized because they have more components that are indeed relevant.  \\

\textbf{2.~BM25.}
Also known as Okapi BM25 (BM stands for Best Matching), this BOW based probabilistic retrieval model ranks documents based on the number of query terms present in each document. 
BM25 ignores inter-relationship between the query terms. 
For a given query $q$ with query terms $q_1,q_2,q_3,...,q_n$, the BM25 relevance score of document D can be calculated as:  

$score(q, D) = \sum_{i=1}^{n} IDF(q_i) (\frac{tf_i (k_1+1)}{tf_i+k_1(1-b+b\frac{|D|}{avg|D|})})(\frac{qtf_i(k_2+1)}{k_2+qtf_i})$.

where, for the query term $q_i$, IDF is the inverse document frequency, and $tf_i$ and $qtf_i$ are term frequencies \wrt to D and q respectively. $|D|$ is the document length and $avg|D|$ is the average document length across all the candidate documents in the corpus. Here, $b$, $k_1$, and $k_2$ are hyper parameters. Notice that, $k_2$ controls the query term frequency and can be ignored for short query like free text search where frequency per query term is usually $1$.
However, in our settings, depending on the SE artifacts, the query can be long and may contain repeated terms that may have important contributions towards the overall similarity measurement. Hence, we keep this $k_2$ hyper parameter in the equation.

{\em Implication:}
A distinguishing feature of BM25 is that it treats a matching term's importance in the document and query differently and also gives a special attention to that term's frequency in the query. 
This characteristic helps BM25 to show improved performance when the query and document are of different types. On the other hand, the document length normalization helps to predict more accurate ranked score where the documents are of various length. Despite these advantages, BM25 is also a keyword matching model and ignores orders of the word in a sentence. Thus BM25 might fit well where document length varies and the order of the tokens in the document is not important.

\textbf{3.~Latent Semantic Indexing (LSI).}
Landauer and Dumais~\cite{landauer1997solution} proposed LSI that is based on the assumption that words with similar meaning will have similar context. LSI projects a higher dimensional document-term co-occurrence frequency matrix into a lower dimensional latent space to create document vectors. 
An effective way of using LSI is to use the  tf-idf weight instead of raw co-occurrence count of term. The  idf  can be estimated from the document corpus. After inferring the lower dimensional vector of both query and candidate documents, cosine similarity can be used to compute the similarity between two document vectors as $sim(q, D)$ (equation above). 

{\em Implication:}
Intuitively, the dimension reduction step computes similarity scores of every word \wrt every other based on their co-existence in a common context.
 In this way, LSI captures the meaning of synonymous 
and polysemy 
words in the latent space. As opposed to VSM and BM25, LSI can better differentiate the documents with synonymous and polysemic words but little semantic similarities. 

\textbf{4.~Word Embedding.} 
Similar words should have similar context~\cite{harris1954distributional}; this observation leads to {\em Word Embedding},  a natural language processing (NLP) technique, where each word $w$ is represented by a d-dimensional vector of real numbers. This d-dimensional vector is learned from the context of $w$ where {\em context} is formed by the preceding and following words of $w$ in a sentence. Similar words should have similar context words thus similar embedding.
Many popular similarity measures like cosine similarity can be used to measure similarities between the embedded documents. Among them {\em Word Mover's Distance (WMD)} is proved to be the winner~\cite{kusner2015word}. 
For each query term, WMD searches for the semantically closest document term in each document, where the distance between two terms is calculated as a Euclidian distance in word embedding space. The summation 
of the minimum distances for all query terms represents the distance from a query to a candidate document.

{\em Implication: }
As word embedding capture the contextual information of words, WMD is useful to bridge the semantic gap between documents. For example, say the descriptions of two projects are ``image gallery app for Lollipop'' and ``Android photo viewer''. They are very close in meaning but have no shared words.  Thus, traditional similarity measures like keyword based BOW model 
could not find any similarity between these two documents. In contrast,  WMD can efficiently judge they are highly similar since they have very similar word embedding. In different SE artifacts, some synonymous terminology is common; {upgrade} and {update} often used interchangeably. WMD might be useful to detect similar documents even if the documents contain no identical words.

\subsection{Studying SE Tasks}
\label{sec:setask}
We analyze the effect of different similarity measure on different types of software document \wrt two IR-based SE tasks:

(i) \textbf{\ProjectTask .} Given a project as a query, this task tries 
to find functionally similar projects from \gh.  A ranked list of projects
is retrieved with the most relevant projects at the top. For example, {screenbird}~\cite{screenbird} and {FFmpegRecorder}~\cite{FFmpegRecorder} both are {Video Recorder} software. For a query with the first project, the system tries to return a list of Video Recorder projects that includes the second project  (see Table~\ref{tab:project-search-example}). 

(ii) \textbf{\BugTask .} Given a bug report as query, this task ranks all the source files in the project repository based on their relevance with the query. The files that top the ranking are more likely to contain the cause of the bug. For example, for bug report id 369884~\cite{bug-report-369884} in Eclipse-\-Platform-UI~\cite{eclipse-ui-link} project, file \textit{E4\-Application.java}~\cite{E4Application-java-file} was fixed  (see Table~\ref{tab:bug-result-example}). 
A perfect \bugtask tool will rank this file at top 
if queried with above bug report.

\section{Methodology}
\label{sec:method}
Here, we describe the dataset and the analysis methods we used to conduct the study. 

\subsection{Study Subject}

We analyze a wide variety of projects for studying two SE tasks that are introduced in Section~\ref{sec:setask}. For \projecttask task, we use a total $1832$ \gh Java projects of $112$ functional categories. Details of the dataset can be found in Table~\ref{tab:github-project-cumulative-data-stats-category}.
For the \bugtask task, we collect a benchmark bug report dataset~\cite{ye2014learning,ye2016word}, which contains $1100$ bug reports from four projects. Details of the dataset are shown in  Table~\ref{tab:study-subject-bug}.

\subsection{Data Collection}
\label{method:project-data-annotation}

For studying \projecttask, we use \gh open source projects. 

\noindent
\textbf{Collecting \gh Projects.}
We use two different methods to select \gh projects. 
First, in (i) \textit{Method-A}, we search \gh with keywords representing project functionalities (\eg media player, text editor, \etc), and download the projects retrieved by the \gh search. In this way we download 1590 projects with 78 different functionalities. As \gh search primarily looks at project descriptions, a project without proper description may have suffered from this step. Hence, in (ii) \textit{Method-B}, we download some \gh projects first and then determine their functionalities.  In the following section, we describe these methods in details. 

\input{study_subj.tex}
  
\textit{Method-A.}
Given a project functionality, we use \gh API to search for the corresponding projects\textemdash project functionalities are represented by the search keywords.  
For example, for retrieving projects related to {\em Video Recorder}  application, we search with keyword {\em Video Recorder}. 

We begin with selecting a meaningful set of project functionalities that we can use as keywords to search projects. We leverage DMOZ Ontology~\cite{dmoz_data_dump}, which is a hierarchical directory of the Web. 
In this ontology, any category under `software' represents a meaningful functionality (\eg Spelling Software, Grammar and Spell Checkers, etc.). We remove the homonyms to reduce confusion of the search task. This approach reduces the category set 
to a size of 90, where each phrase represents a certain software functionality, such as Spell Checker. 

We use these categories as queries to search \gh for relevant Java projects using \gh search API~\cite{githubsearchapi}. 
We exclude the forked projects as they include near-identical projects and overfit our project similarity data. 
For each query, we select top 1,000 projects from the search result. Further, we only select projects with ratings 3-star 
and above to focus on important and (hopefully) well documented projects. Thus, we end up having $2180$ unique projects under 90 project categories, where some projects may belong to multiple categories.  

We further manually investigate the associated categories of each project, because \gh search is mostly based on keyword matching and in some cases it leads to inaccurate categorization. For example, projects \textit{Eid-Applet}~\cite{misclass-project1}'s description is ``{eID Applet to enable BE eID cards within web browsers} and it is retrieved by the query {Web Browser}''. The retrieved project is certainly not a Web Browser but an Applet. While manually investigating the project annotation, we further modify, delete, and add categories (\ie functionalities) as needed. We also remove some ambiguous projects. 
After such rigorous filtration process, we end up with \simprojdata projects under \simprodatacategory different functionalities. Table~\ref{tab:github-project-cumulative-data-stats-category} presents the details.

Notice that, the collected projects are popular (above 3-star rating) and resulted from a keyword based search. Thus, most of the projects contain proper description, readme, and other source code content. However, there are many projects on \gh that do not have adequate textual artifacts (\DES, \RME, etc.). To study well-represented diverse projects, we further enhance the data set using the following method. 

\textit{Method-B.}
First, we collect $242$ \gh projects that have Google play links in their descriptions or README contents. Note that we exclude forked projects and consider popular project having at least $3$-start to remove any potential toy project~\cite{kalliamvakou2014promises}. Then we manually annotate them using the details available in the Google Store. 
We leverage the app description, similar app suggestion, category, etc. information available in Google Play Store to annotate these projects. In this way, we can annotate a \gh project with its functionality even if it does not contain elaborate textual artifacts except Google play link (in \DES or \RME). We also find some new project functionalities that have not seen in Method-A.  
Finally, we find 1832 ($1590+242$) projects 
with 112 different functionalities (see Table~\ref{tab:github-project-cumulative-data-stats-category}), where  
Media Player, Search Engine, Database Systems, \etc are top functionalities with maximum member projects. 

For both methods, two researchers annotated project category {\em separately} and resolved the disagreements by discussion. We observe that $95\%$ of the cases they agreed on project categories. Note that, as the annotator needs to consult various documents (\ie description, readme, Google store, etc) to come up with a functional category, it required tremendous manual effort to annotate all the projects. Per annotator, it took on average $3$ minutes per project and in total approximately $90$ working hours to finish all the annotations. We will open-source our dataset to encourage further research.

\noindent
\textbf{Collecting Bug Report Data.}
We collect a benchmark bug report dataset and study four projects' data: Birt~\cite{birt-link}, Eclipse Platform UI (Eclipse-UI)~\cite{eclipse-ui-link}, Eclipse JDT~\cite{eclipse-jdt-link}, and SWT~\cite{swt-link}. This dataset has been used previously for the \bugtask  task~\cite{ye2016word,ye2014learning,ye2016mapping,lam2017bug}. In this dataset, each bug report contains a summary, description, report time, and status of its fix along with bug fix commit. For each bug report, using the bugfix commit, we download the before-fix version of the project. 
The files that are added, deleted, and modified in the bugfix commit are considered as the true buggy files. As the added files are not part of the before-fix version, the system cannot predict it, so we ignore that for the evaluation. Table \ref{tab:study-subject-bug} shows details of the report dataset.

\subsection{SE Artifacts Extraction}
\label{sec:fc}
We extract following documents from our dataset \wrt two SE tasks.

\noindent
\textbf{Features of \projecttask task}. From the collected {project data}, we choose five types of SE artifacts:

\noindent
(i) {Project Description:} This textual artifact is often short and concisely represent the project functionality. 

\noindent
(ii) {Readme Content:} Textual artifact, usually contains a detailed description including how to install and run the project. 

\noindent
(iii) {Method \& Class name:} Developers often use meaningful identifier names to implement project~\cite{allamanis:fse:14}. Thus, it might be possible that projects with similar functionalities use similar method or class names. 
For example, two text editor applications may have similar methods with names {copy}, {paste}, {save} etc. 
To check this hypothesis, we retrieve method and class names that are declared within a project. 

\noindent
(iv) {Import Package name:} Similar projects often use similar API packages~\cite{mcmillan2012detecting}.
This motivates us to use imported API package names and class names as features. We use Eclipse JDT~\cite{eclipse-jdt-link} framework to collect these names. 

\noindent
(v) {API name:} The API Class refers to the classes defined in system libraries or other third-party libraries or packages. To extract these, using Eclipse JDT we first extract all the classes used in a project and then remove the classes defined within the project from this list. The remaining class names are considered as API names.

\noindent
\textbf{Features of \bugtask task}.
For each project we consider the source files, which mostly contain code tokens, as candidate documents to compute their similarity with the bug report query. 

\subsection{Data Pre-processing}
\label{sec:data-preprocessing-project}

For each feature, we use standard natural language processing (NLP) techniques for data processing like tokenization, normalization, stemming, and stopword removal. 
First, we clean the documents by removing the special character (\ie non-English) and punctuation from it. As a convention, Java usages camel case format for its class, method or any variable name.  For such compound tokens (\ie \textit{TerminalFactory}), both in textual and code artifacts, we further extract smaller token units (\ie \textit{Terminal} and \textit{Factory}). We also keep the original compound token in code artifacts to keep actual keyword information. We then normalize the tokens: remove the numeric character and convert to lower case letter. To avoid bias from the frequently occurring but less informative tokens we remove two types of stopword: standard English stopword list (adopted from ~\cite{english-stopword-link}) and Java language related stopwords, that is keywords~\cite{java-keyword-link}: \textit{void}, \textit{public}, \textit{while}, etc.
 To reduce the unwanted lexical gap between tokens, we apply Porter Stemmer\cite{porter1980algorithm} to convert words to its base form (\ie convert \textit{computes} and \textit{computed} into \textit{comput}).

\subsection{Evaluation Metric}

We evaluate an IR task \wrt its ground truth sets, \ie, given a query and a candidate document, we check whether the retrieved results are matched with its corresponding ground truth.  We use several standard evaluation metrics~\cite{Manning:2008:IIR:1394399} as described below:

\noindent
\textit{1.~Precision (P).} For a given query $q$, precision is the fraction of retrieved documents that are also present in the ground truth set. Thus, $P = \frac{r}{d}$, where $r$ is the number of relevant items from the retrieved $d$ documents.

\noindent
\textit{2.~Recall (R).} For a given query $q$, recall is the fraction of relevant documents that are retrieved. If $t$ be the total relevant documents for the query $q$, the recall is $R = \frac{r}{t}$.




\noindent
\textit{3.~Mean Average Precision (MAP).}
For a set of queries, $MAP$ is the mean of the average precision of  individual query~\cite{Manning:2008:IIR:1394399}. 
First, for each query, an average precision is computed for each rank. Given a query($q$) and it's ranking documents, average precision of $q$ is calculated as $AvgPrec(q) = (\sum_{i=1}^{R} \frac{i}{rank_i})/R$, where $R$ is the total number of relevant documents,
$rank_i$ is a ranking position of the relevant document $i$ in the retrieved ranking and $i/rank_i= 0$ if relevant document $i$ was not retrieved by the model.
Then we take the mean of this average precision across all the queries using equation $MAP(Q) = \frac{1}{|Q|}\sum_{j=1}^{|Q|}AvgPrec(q_j)$ to get $MAP$. Here, $Q$ is the entire query set. 

\noindent
\textit{4.~Mean Reciprocal Rank (MRR).}
Given a retrieved list for a query, the reciprocal rank is computed as the multiplicative inverse of the rank of the first relevant document. Mean of such reciprocal rank across all the queries are taken using equation  $MRR(Q)=\frac{1}{|Q|}\sum_{i=1}^{|Q|}{\frac{1}{rank_{i}}}$. Here, $rank_{i}$ is the rank position of the first relevant document for the i$^{th}$ query.


We evaluate a search result by computing these evaluation metrics at different rank cut-off. During comparison we use percentage gain computed as $gain = (b-a)/a*100$, any metric value changes from $a$ to $b$.

\subsection{Model Configurations}
Performance of IR models varies significantly with different parameter settings~\cite{biggers2014configuring,panichella2013effectively}. For a fair comparison, we tune all the models to its best performing configuration for each task, as shown in Table ~\ref{tab:model-best-config}. 
We did an exhaustive search\textemdash we varied the parameter values at regular intervals and then chose the optimal ones. \todo{give exact interval number and some sample results}
\input{table-model-config}


We train a skip-gram based word2vec~\cite{mikolov2013distributed} word embedding model  which is used by WMD. We use a diverse collection of $3.7M$ Wikipedia articles~\cite{wikidump} and $7.5M$ \gh projects description collected using GHTorrent \cite{Gousi13} for the training data. 
We use Gensim's~\cite{rehurek_lrec} Python implementation of word2vec to train on our data. 
For WMD in \bugtask task, we use a pre-trained word2vec model which is trained on the data containing source code and API documentation and found to be effective on the same dataset for \bugtask~\cite{ye2016word}. 

\section{Case Studies}
\label{sec:case-study}

Our central question is whether a choice of similarity measure matters while computing similarities between different types of SE artifacts, especially for IR related SE tasks.  Thanks to the software forges like \gh, Bitbucket, Bugzilla \etc various types of textual and code related SE artifacts are available.  
First,  we collect a ground truth set of closely matched documents corresponding to the two studied tasks. Next, we check how different similarity measures perform to find relevant documents \wrt to this ground truth set. 
Inspired by our empirical results, we investigate whether such an informed choice of similarity measure impact the performance of IR based SE tools. We analyze two previously proposed tools for \projecttask and \bugtask tasks.

 \subsection{Case 1: \ProjectTask}
\label{sec:case-study-project}

In \projecttask task, for a given project set $N$, we take one project as the query and consider rest $N-1$ projects as candidate documents. If the categories of a retrieved project and the query project are identical, we consider that as a success.  We take the average performance across all the projects in a randomly chosen query set of $200$ to get overall performances of different models. We discuss the results mostly \wrt $MAP@10$. However, a similar conclusion can be drawn from all other evaluation metrics.

\noindent
\textbf{Text-Text Artifact.}
We study the impact of various similarity measure while computing similarities between textual documents, \eg description vs.~description, and readme vs.~readme. Project recommendation task is a classic example of this. Given a project name as a query, the query string is augmented with textual feature documents. The candidate projects are then searched using different similarity measure. 
We first investigate: 

\RQA{1}{How different similarity measures perform for Text-Text artifacts?}

\begin{table*}[t]
  \vspace{0.3cm}
  \centering
  \caption{ {Different models performance on \textbf{textual artifacts} for \projecttask task. Best performing  values marked in \textcolor[rgb]{ 1,  0,  0}{Red} (\textbf{bold}).}}
    \resizebox{0.95\textwidth}{!}{%

    \begin{tabular}{l|cccc|cccc}
    \toprule
          & \multicolumn{4}{c|}{Description} & \multicolumn{4}{c}{Readme} \\
          & VSM   & LSI   & BM25  & WMD   & VSM   & LSI   & BM25  & WMD \\
    \midrule
    MAP@10 & 0.51  & \textcolor[rgb]{ 1,  0,  0}{\textbf{0.57}} & 0.51  & 0.51  & 0.37  & \textcolor[rgb]{ 1,  0,  0}{\textbf{0.39}} & 0.26  & 0.29 \\
    MRR   & 0.56  & \textcolor[rgb]{ 1,  0,  0}{\textbf{0.61}} & 0.55  & 0.57  & 0.44  & \textcolor[rgb]{ 1,  0,  0}{\textbf{0.45}} & 0.30  & 0.34 \\
    P@10  & 0.40  & \textcolor[rgb]{ 1,  0,  0}{\textbf{0.49}} & 0.41  & 0.33  & 0.25  & \textcolor[rgb]{ 1,  0,  0}{\textbf{0.29}} & 0.14  & 0.14 \\
    R@10  & 0.13  & \textcolor[rgb]{ 1,  0,  0}{\textbf{0.16}} & 0.13  & 0.10  & 0.07  & \textcolor[rgb]{ 1,  0,  0}{\textbf{0.08}} & 0.03  & 0.03 \\
    \bottomrule
   
    \end{tabular}%
    }
     \vspace{0.3cm}
  \label{tab:textual-feature-results}%
\end{table*}%

In Table \ref{tab:textual-feature-results}, we see that for the textual artifacts, in general, LSI achieves the highest performance for most of the evaluation settings. For \textit{project description}, LSI model performs best for all the evaluation metrics and gains $11.76\%$ compared to the second best models (WMD, VSM, and BM25) at $MAP@10$.
In our dataset, WMD, VSM, and BM25 perform comparably. 
Similarly, for \textit{readme}, LSI model performs best in most of the evaluation metrics \textemdash 4.85\% improvement for $MAP@10$ \wrt VSM, the second best model. 

Table~\ref{tab:project-search-example} shows example ranked lists for all four models using \textbf{\DES} only feature for a query project `screenbird' with Video Recorders category. The top two projects retrieved by LSI and VSM models are of same categories as the query. Notice that all these retrieved projects have keyword `video', and `recording' in their descriptions.  However, VSM mistakenly retrieves a media player app as the third project because of the word `video' presents in the document description.

\begin{table}[t]
\setlength\tabcolsep{2pt}
\vspace{0.3cm}
    \centering
  \caption{{Top three Search Results for \projecttask task with \textbf{\DES} only  feature. 
  The keywords in \textbf{bold} highlight the important keywords  for matching.}}
    \vspace{0.1cm}
    \begin{tabular}{lp{30em}}
    \toprule
    Category & Name : Description  \\
    \midrule
     \underline{\bf Query} &  \\
     \textbf{Video Recorder} & screenbird :a full \textbf{cross} \textbf{platform} \textbf{video} \textbf{screen} capture tool and \textbf{host}. Java based \textbf{screen} \textbf{recorder}, and Django based \textbf{web} backend distributed \textbf{video} processing \textbf{engine} that uses \textbf{ffmpeg} ... of AWS instances. \\
    \midrule
    \midrule
     \underline{\bf LSI} &  \\
    (1) Video Recorder & FFmpegRecorder : An Android \textbf{video} \textbf{recorder} using ... \textbf{FFmpeg}. \\
    (2) Video Recorder & FFmpegVideoRecorder : Customizable Android \textbf{video} \textbf{recorder} library... \\
    (3) Video Recorder & jirecon : A Standalone \textbf{recording} container for  ... \textbf{video} \textbf{recorder} ... \\
     \midrule
     \underline{\bf VSM} &  \\
    (1) Video Recorder & ScreenRecorder : containing service for \textbf{recording} \textbf{video} of device screen \\
    (2) Video Recorder & VideoRecorder : Android \textbf{video} \textbf{recorder} project \\
    (3) Media Player & dttv-android : android \textbf{video} player based on dtplayer \\
     \midrule
     \underline{\bf BM25} &  \\
  (1) Video Recorder & ScreenRecorder : containing service for \textbf{recording} \textbf{video} of device screen \\
  (2) Terminal Emulator & DragonConsole : a \textbf{cross} \textbf{platform} Java based Terminal Emulator. \\
  (3) Search Engine & LunarBase : real-time engine, ... \textbf{records} in one table, ... used as a search \textbf{engine}\\
    \midrule
    \underline{\bf WMD} &  \\
  (1) Media Player & supersonic : \textbf{web}-based \textbf{media} \textbf{streamer} ... \textbf{audio} and \textbf{video} formats \\
  (2) Readers Java & java-manga-reader : directly from \textbf{web}. .., \textbf{internet} access is required \\
   (3) Search Engines & SearchEngine :  crawls seed \textbf{web} page ... search \textbf{engine} for a \textbf{website} ... \\
 \bottomrule
    \end{tabular}%
  \label{tab:project-search-example}%
  \vspace{0.3cm}
\end{table}%

Interestingly, all the projects retrieved by WMD in Table~\ref{tab:project-search-example} are wrong, and they have minimum keyword overlapping with the query projects. However, a closer look will reveal the document and query terms are related. This indicates projects description of \gh is not very complex. We may not need a word embedding based similarity measure where contextual similarity pays a pivotal role and may undermine simple keyword based matching. 

\RS{1}{Context-aware model LSI performs the best for text-text documents.}

\noindent
\textbf{Code-Code Artifact.}
Similar to textual artifacts, we study the impact of various similarity measures while computing similarities between code documents, \eg method-class vs.~method-class, and API vs.~API, etc. We leverage document features from \projecttask for this RQ. Next, we investigate: 

\RQA{2}{How different similarity measures perform for Code-Code artifacts?}
 While the performance of LSI is higher in textual artifacts, VSM model starts dominating in performance for code artifacts. Compare to LSI, VSM performs slightly better at \textit{method and class} feature and significantly better for \textit{import package} and \textit{API} names with $20.59\%$ and $24.29\%$ improvement respectively at $MAP@10$.  

\begin{table*}[t]
  \vspace{0.3cm}
  \centering
  \caption{ Different models performance on \textbf{code artifacts} for \projecttask task. Best performing  values marked in \textcolor[rgb]{ 1,  0,  0}{Red} (\textbf{bold}).}
  \vspace{0.1cm}
    \resizebox{0.95\textwidth}{!}{%
    
        \begin{tabular}{l|cccc|cccc|cccc}
    \toprule
          & \multicolumn{4}{c|}{Method Class } & \multicolumn{4}{c|}{Import Package} & \multicolumn{4}{c}{API} \\
          & VSM   & LSI   & BM25  & WMD   & VSM   & LSI   & BM25  & WMD   & VSM   & LSI   & BM25  & WMD \\
    \midrule
    MAP@10 & \textcolor[rgb]{ 1,  0,  0}{\textbf{0.37}} & 0.36  & 0.16  & 0.25  & \textcolor[rgb]{ 1,  0,  0}{\textbf{0.29}} & 0.24  & 0.07  & 0.20  & \textcolor[rgb]{ 1,  0,  0}{\textbf{0.31}} & 0.25  & 0.22  & 0.25 \\
    MRR   & \textcolor[rgb]{ 1,  0,  0}{\textbf{0.43}} & 0.40  & 0.19  & 0.29  & \textcolor[rgb]{ 1,  0,  0}{\textbf{0.34}} & 0.29  & 0.08  & 0.26  & \textcolor[rgb]{ 1,  0,  0}{\textbf{0.35}} & 0.28  & 0.26  & 0.30 \\
    P@10  & 0.23  & \textcolor[rgb]{ 1,  0,  0}{\textbf{0.24}} & 0.07  & 0.10  & \textcolor[rgb]{ 1,  0,  0}{\textbf{0.16}} & 0.13  & 0.03  & 0.08  & \textcolor[rgb]{ 1,  0,  0}{\textbf{0.17}} & 0.13  & 0.10  & 0.12 \\
    R@10  & 0.06  & \textcolor[rgb]{ 1,  0,  0}{\textbf{0.07}} & 0.02  & 0.03  & \textcolor[rgb]{ 1,  0,  0}{\textbf{0.05}} & 0.03  & 0.01  & 0.02  & \textcolor[rgb]{ 1,  0,  0}{\textbf{0.05}} & 0.04  & 0.04  & 0.04 \\
    \bottomrule
    \end{tabular}%
    }
    \vspace{0.3cm}
  \label{tab:feature-results}%
\end{table*}%
Among code features, method and class names perform best. This  suggests that similar projects actually have similar method and class names. This finding also confirms the hypothesis of Allamanis \etal~\cite{allamanis:fse:14} that developers use meaningful identifier names while writing a software program.

Interestingly, both the context aware models LSI and WMD show a similar decreasing trend of performance going from textual artifacts to code artifacts 
(see Table~\ref{tab:feature-results}). This result indicates that the contextual feature, which is quite effective in natural language text, is not that helpful for source code artifact.
Further, among source code artifacts, LSI and WMD perform better for method class names suggesting context is more meaningful for these names than import packages and API names.

On the other hand, BM25 performs significantly worse for the rest of the artifacts.  BM25 treats query and document differently (see the scoring equation in Section \ref{sec:back}), although for homogeneous artifacts (\eg code vs. code and text vs. text) the query and documents are linguistically identical. In addition, BM25 assumes all the terms in the query are important.
For example, in Table~\ref{tab:project-search-example}, except for the top project, the rest are not correct for BM25. As it ignores inter-relationship of query terms, it emphasizes on all parts of the query and thus misguided by the variety of concept in the query. Thus, a verbose query might hurt the performance of BM25. 

\RS{2}{Keyword based bag-of-words model VSM performs best for code only artifacts. Code artifacts lack context information.}
\subsubsection{Implication}
Previously proposed tool, CLAN~\cite{mcmillan2012detecting} compares JDK APIs (packages and classes) used in the studied projects using LSI algorithm to establish similarities. 
Since CLAN's source code is not available, we reimplement CLAN adhering to the paper details. 
We further extend CLAN to incorporate all the APIs studied in our dataset including JDK APIs. 

We find that VSM is the best performing measure for the artifacts type used by CLAN (see RQ2). Thus, we build a modified version of CLAN, named vsmCLAN, where we replace the similarity model used by CLAN with VSM.  
CLAN usages a weighted combination of both features' similarity value for the final score. Thus, we exhaustively tune the weights in an interval of $0.1$ and report optimal performance for the comparison. 
For this experiment we consider a query set consisting of $1590$ projects (\ie method-A in Table~\ref{tab:github-project-cumulative-data-stats-category}) and report average performance.
We observe that original CLAN achieve the best results at weights \textit{import package} $=0.9$ and \textit{API} $=0.1$. and modified vsmCLAN achieve the best results at weight \textit{import package} $=0.6$ and \textit{API} $=0.4$. Next, we analyze this research question.

\RQA{3}{Can an informed choice of IR model improve the performance of \projecttask task?}
\label{rq3}

\begin{figure}[ht]
\centering
\includegraphics[width=0.5\textwidth]{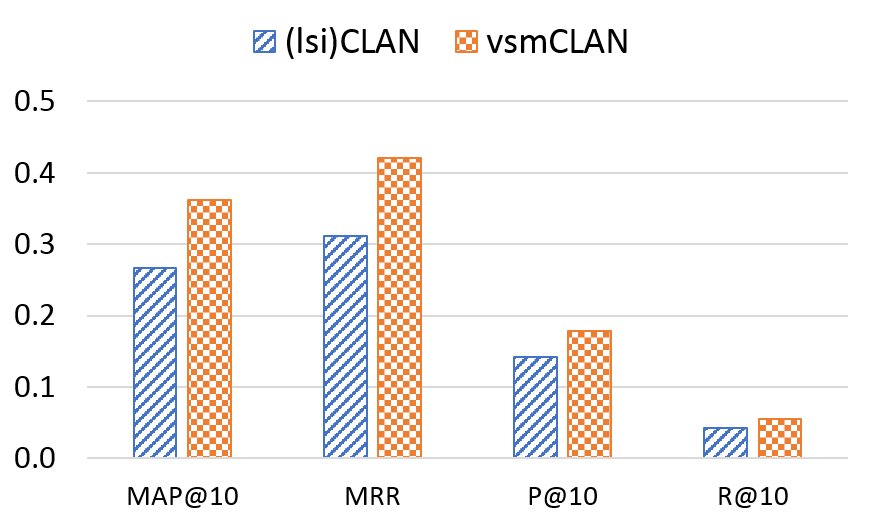}
\caption{\textbf{\small Comparison with CLAN}
}
\label{fig:clan-case-study}
\vspace{0.5cm}
\end{figure}

Figure~\ref{fig:clan-case-study} shows that our modified version vsmCLAN outperforms the original tools (lsi)CLAN by $35\%$ at MAP@10, $35\%$ at MRR, $25\%$ at P@10, and $29\%$ at R@10. Though all the system architecture remains the same for the modified version, better performance of VSM on code artifacts (\ie package and API) contributed to the overall improved performance of the tool.

All these results strongly suggest that with a right choice of similarity measure can really improve the performance of the existing tool. 

\RS{3}{A recommendation tool built with an informed choice of similarity measure can significantly improve the performance of original tool.}

\subsection{Case 2: \BugTask Task}
 We empirically evaluate the performance of various similarity measures given the bug reports as a query. Here, we only consider {source code}  as the documents.

\noindent
\textbf{Mixture Artifact.}
In this case, the query and candidate document types are different (text vs.~ code). We check the similarities between bug reports and the source files by studying $1100$ bug reports from four projects (Table~\ref{tab:study-subject-bug}). If the model ranks the actual buggy file at the top, we consider that as a success. Next, we question:


\RQA{4}{How different similarity measures perform across a mixture of text and code artifacts?}
\label{rq4}


\begin{table}[t]
\vspace{0.3cm}
    \centering
  \caption{ {Different models performance on\textbf{ mixture artifacts} for \bugtask. Best performing models' values marked as
\textcolor[rgb]{ 1,  0,  0}{Red} (\textbf{bold}) for each project and evaluation metric.
}}
    \resizebox{0.95\textwidth}{!}{%
    \begin{tabular}{l|cccc|cccc|cccc|cccc}
    \toprule
          & \multicolumn{4}{c|}{Birt}     & \multicolumn{4}{c|}{Eclipse-UI} & \multicolumn{4}{c|}{JDT}      & \multicolumn{4}{c}{SWT} \\
          & VSM   & LSI   & BM25  & WMD   & VSM   & LSI   & BM25  & WMD   & VSM   & LSI   & BM25  & WMD   & VSM   & LSI   & BM25  & WMD \\
    \midrule
    MAP@10 & 0.11  & 0.03  & \textcolor[rgb]{ 1,  0,  0}{\textbf{0.17}} & 0.02  & 0.10  & 0.06  & \textcolor[rgb]{ 1,  0,  0}{\textbf{0.29}} & 0.02  & 0.06  & 0.02  & \textcolor[rgb]{ 1,  0,  0}{\textbf{0.28}} & 0.00  & 0.10  & 0.11  & \textcolor[rgb]{ 1,  0,  0}{\textbf{0.42}} & 0.01 \\
    MRR   & 0.13  & 0.04  & \textcolor[rgb]{ 1,  0,  0}{\textbf{0.18}} & 0.02  & 0.12  & 0.08  & \textcolor[rgb]{ 1,  0,  0}{\textbf{0.30}} & 0.02  & 0.08  & 0.03  & \textcolor[rgb]{ 1,  0,  0}{\textbf{0.31}} & 0.01  & 0.12  & 0.13  & \textcolor[rgb]{ 1,  0,  0}{\textbf{0.44}} & 0.01 \\
    P@10  & 0.03  & 0.02  & \textcolor[rgb]{ 1,  0,  0}{\textbf{0.05}} & 0.02  & 0.04  & 0.03  & \textcolor[rgb]{ 1,  0,  0}{\textbf{0.08}} & 0.02  & 0.02  & 0.01  & \textcolor[rgb]{ 1,  0,  0}{\textbf{0.07}} & 0.00  & 0.04  & 0.04  & \textcolor[rgb]{ 1,  0,  0}{\textbf{0.10}} & 0.01 \\
    R@10  & 0.13  & 0.05  & \textcolor[rgb]{ 1,  0,  0}{\textbf{0.23}} & 0.02  & 0.16  & 0.08  & \textcolor[rgb]{ 1,  0,  0}{\textbf{0.45}} & 0.02  & 0.14  & 0.04  & \textcolor[rgb]{ 1,  0,  0}{\textbf{0.43}} & 0.00  & 0.18  & 0.20  & \textcolor[rgb]{ 1,  0,  0}{\textbf{0.55}} & 0.01 \\
    \bottomrule
    \end{tabular}%
    }
    
  \label{tab:bug-model-result}%
  \vspace{0.3cm}
\end{table}%



We observe that BM25 is the best performing model across all the projects, as shown in Table ~\ref{tab:bug-model-result}. BM25 outperforms other models significantly and achieves a percentage gain of (MAP@10, MRR): Birt ($54\%$, $38\%$), Eclipse-UI ($190\%$, $150\%$), JDT ($366\%$, $287\%$), and SWT ($320\%$, $266\%$), compared to the second best VSM model. Table~\ref{tab:bug-result-example} shows an example, where the BM25 ranks the intended file at top 1 position, where VSM ranks it at 31$^{st}$ places. 
\vspace{0.3cm}
\input{table-bug-localization-example-results}
\vspace{0.3cm}

Here bug report is human written summary of a bug of software thus contains important keyword to locate bug in a software project. Thus, bug reports are  usually short in length and free of unnecessary repetition of the term. On the other hand, the source code document contains mostly code token. Though the variation of length of different source file might be responsible for the worst performance of the other models, BM25 mitigate its impact with the document length normalization factor. Thus the nature of query (bug report) and document (source code) makes BM25 a better choice among all the models, which is also reflected by the evaluation results. 

VSM and LSI assume both query and document are of same type and represent them in the same concept space. This assumption works better for homogeneous artifacts: text-text in RQ1 and code-code in RQ2. But for mixture artifacts like bug reports vs.~ source code, representing them in the same concept space is less effective and that is also indicated by the results in Table ~\ref{tab:bug-model-result}. 

Similar to RQ2, we also see that contextual feature in source code is less effective. Due to the the co-existence of text and code in bug reports, the context feature is less in the bug reports as well. This lack of context in the query and document might be a reason for the poor performance of context aware models (LSI and WMD). 

\RS{4}{For mixture (text-code) documents, BM25 performs the best. Surprisingly, BM25 is not that effective for text only and code onlyartifacts.}

\subsubsection{Implication}
Ye \etal~\cite{ye2014learning} proposed a learning to rank based bug localization tool (LR) which leverages the VSM similarity measure.
We implement LR tool adhering the paper details. LR combines six feature score to get the final score for a source file corresponds to a bug report query. Among six features, three features: Class Name, Bug fix time, bug fix frequency are meta information. We follow the same technique as Ye \etal ~\cite{ye2014learning} to compute these feature values for both the tools. Please refer to the Ye \etal ~\cite{ye2014learning} for feature extraction details.
Ground truth collection from benchmark bug-report dataset can be found in Section~\ref{sec:method}. 

Our experiments show that BM25 performs best for such a mixture of textual and code artifacts. Thus we replace similarity measure of LR tool from VSM to BM25 to build bm25-LR as in Table~\ref{tab:lr-case-study-model-assign}. Note that meta-features do not require any similarity measure, thus both original (vsm)LR tool and modified bm25-LR uses the same value for these features. For this experiment we consider the latest $100$ bug reports for each of the four projects (in Table~\ref{tab:study-subject-bug}) from our dataset. 
Note that LR model leverages a learning to rank to find out optimal weights to combine features~\cite{ye2014learning}. Thus the learning weights heavily depend on the training data. To mitigate such unwanted bias, we apply an exhaustive search by varying weights in regular interval (\ie 0.05) to achieve optimal parameter settings. The optimal feature weights for (vsm)LR and bm25-LR are given in Table~\ref{tab:weights-lr} and Table~\ref{tab:weights-bm25lr} respectively. Then, we investigate the following research question.

\begin{table}[t]
  \vspace{0.3cm}
  \centering
  \caption{Model assignment of features for \bugtask task}
   \resizebox{0.95\textwidth}{!}{%
   \begin{tabular}{l|cccccc}
    \toprule
     Feature & Source & API   & Collab.   & Class & Bug Fix & Bug Fix\\
     & Code  & Descr. & Filter & Name  & Time  & Freq\\
    \midrule
    bm25-LR & BM25  & BM25  & BM25  & meta-value    & meta-value   & meta-value \\
    (vsm)LR & VSM   & VSM   & VSM   & meta-value   & meta-value   & meta-value\\
    \bottomrule
    \end{tabular}%
  \label{tab:lr-case-study-model-assign}%
  }
  \vspace{0.3cm}
\end{table}%

\begin{table}[t]
\vspace{0.3cm}
  \centering
  
  \caption{Optimal Feature weights for original \textbf{(vsm)LR} \bugtask tool}
  \resizebox{0.95\textwidth}{!}{%
    \begin{tabular}{ccccccc}
          & Source  & API   & Collab   & Class  & Bug-fix & Bug-fix \\
    Project & Code (\textbf{w1}) & Des. (\textbf{w2}) & Filter (\textbf{w3}) &  Name (\textbf{w4}) &  Recency (\textbf{w5}) &  Freq. (\textbf{w6}) \\
    \midrule
    Birt  & 5     & 0.5   & 5.5   & 5.5   & 1.5   & 0.55 \\
    Eclipse-UI & 9.5   & 0.95  & 4.5   & 6.5   & 1.05  & 0.75 \\
    JDT   & 4.5   & 2.6   & 5.5   & 6.5   & 1.05  & 0.55 \\
    SWT   & 4.2   & 3.5   & 4.7   & 7.9   & 0.05  & 0.95 \\
    \bottomrule
    \end{tabular}%
    \label{tab:weights-lr}%
    }
    \vspace{0.3cm}
  
\end{table}%

\begin{table}[t]
\vspace{0.3cm}
  \centering
  \caption{Optimal Feature weights for modified \textbf{bm25-LR} \bugtask tool}
   \resizebox{0.95\textwidth}{!}{%
    \begin{tabular}{ccccccc}
          & Source  & API   & Collab   & Class  & Bug-fix & Bug-fix \\
    Project & Code (\textbf{w1}) & Des. (\textbf{w2}) & Filter (\textbf{w3}) &  Name (\textbf{w4}) &  Recency (\textbf{w5}) &  Freq. (\textbf{w6}) \\
    \midrule
    Birt  & 2.4   & 0.05  & 3.5   & 2.5   & 0.6   & 0 \\
    Eclipse-UI & 3.4   & 0.05  & 3     & 2.5   & 0.5   & 0 \\
    JDT   & 1.2   & 0.25  & 2.5   & 1.2   & 0.3   & 0 \\
    SWT   & 4.4   & 0     & 3.5   & 2.5   & 0.3   & 0.5 \\
    \bottomrule
    \end{tabular}%
     \label{tab:weights-bm25lr}%
    }
\vspace{0.3cm}
 
\end{table}%

\begin{figure}[t]
  \centering
  \begin{minipage}[b]{0.48\textwidth}
    \includegraphics[width=\textwidth]{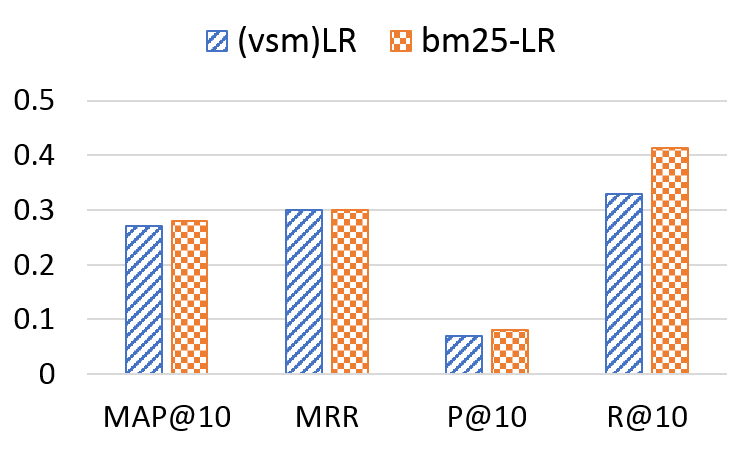}
    \caption{\small Comparison with LR (\textbf{Birt})}
    \label{fig:birt-lr-vs-bm25lr}
    \vspace{0.9cm}
  \end{minipage}
  \hfill
  \begin{minipage}[b]{0.48\textwidth}
    \includegraphics[width=\textwidth]{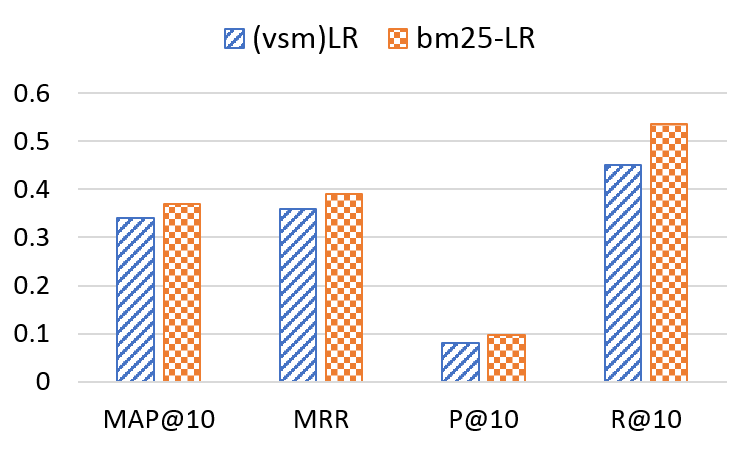}
    \caption{\small Comparison with LR (\textbf{Eclipse-UI})}
    \label{fig:eclipse-lr-vs-bm25lr}
    \vspace{0.9cm}
  \end{minipage}
  
    \begin{minipage}[b]{0.48\textwidth}
    \includegraphics[width=\textwidth]{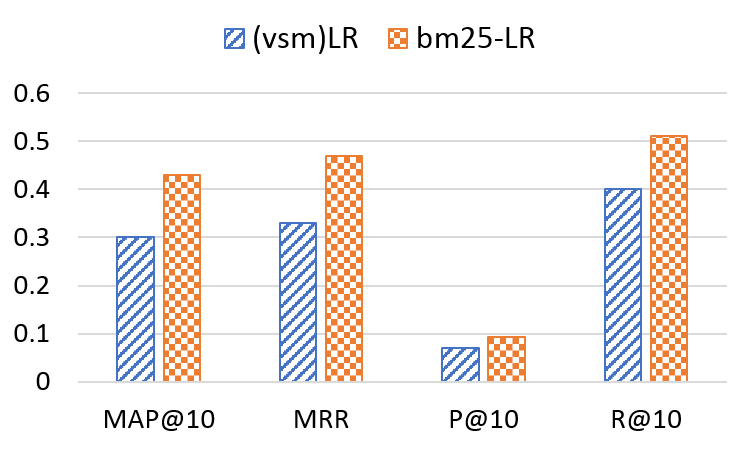}
    \caption{\small Comparison with LR (\textbf{JDT})}
    \label{fig:jdt-lr-vs-bm25lr}
  \end{minipage}
  \hfill
  \begin{minipage}[b]{0.48\textwidth}
    \includegraphics[width=\textwidth]{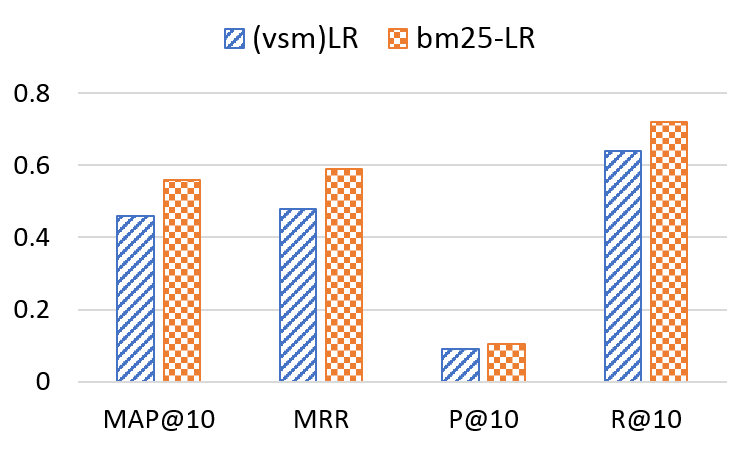}
    \caption{\small Comparison with LR (\textbf{SWT})}
    \label{fig:swt-lr-vs-bm25lr}
  \end{minipage}
  \vspace{0.9cm}
\end{figure}

\RQA{5}{Can an informed choice of IR model improve the performance of \bugtask task?}
\label{rq5}
We observe that for all the projects our modification indeed helps to improve the overall performance in all the evaluation metrics. In JDT, SWT, and Eclipse-UI projects modified bm25-LR achieves a percentage performance gain of $43\%$, $21\%$, and $8\%$ at metric MAP@10 (Figure~\ref{fig:jdt-lr-vs-bm25lr},~\ref{fig:swt-lr-vs-bm25lr}, and~\ref{fig:eclipse-lr-vs-bm25lr}) respectively. 
On the other hand, we find that both the tools perform comparably for the Birt projects (Figure~\ref{fig:birt-lr-vs-bm25lr}). For this project, we explore the feature weights and observe that the overall tool's performance heavily depends on the non-source code features (\eg Class name, bug fix history, etc). Thus the superior performance of BM25 over VSM for source code similarity (as in RQ4) is less visible in the combined score. Similar results also observed previously by Ye \etal~\cite{ye2014learning} on the same dataset.
A similar conclusion can be drawn for other evaluation metrics.

These results indicate that with an informed choice of similarity measure a better result can be achieved for the \bugtask task.
\RS{5}{A \bugtask tool built with an informed choice of similarity measure can significantly improve the performance of original tool.}

\section{Discussion}
\label{sec:discussion}
Our empirical study shows that the performance of different IR models varies with document types. Thus an informed model choice based on document type might help to provide a systematic way to take advantage of all the existing standard similarity measure for a better tool performance. For instance, it has been observed that code search engines perform poorly to satisfy user's code related information need~\cite{hucka2016software,sim2011well}. The inherent problem with such system is that to answer any code query, the search engine has to consult with a diverse set of documents (\eg API documentation, StackOverflow post, \gh issue, etc.). Thus, relying on a standard IR model might result in a very poor overall performance. In contrast, a systematic way of combining different optimal similarity measures might lead to a better search performance. Exploring this in details can be an interesting future work.

\section{Related Work}
\label{sec:related}


Panichella \etal~\cite{panichella2016parameterizing} propose 
a Genetic Algorithm based approach to automatically configure and assemble IR models.
Oliveto \etal~\cite{oliveto2010equivalence} empirically show the equivalence of different similarity measure for traceability recovery task. 
Gethers \etal~\cite{gethers2011integrating} propose an integrated approach to combine orthogonal IR techniques: VSM, probabilistic Jensen and Shannon (JS) model~\cite{abadi2008traceability}, and Relational Topic Modeling (RTM)~\cite{chang2010hierarchical,gethers2011integrating} for traceability recovery task. Evaluating on one repository (EasyClinic) containing $37$ target/candidate documents, they analyze the impact of artifact types (\ie  use cases, UML diagrams, and test cases) on these IR models (\ie JS, JS+RTM, VSM and RTM+VSM). They find that the combination with RTM is highly valuable when tracing with UML diagrams artifact. 


Other parameters may also influence the performance of SE tasks. For example, incorporating user interaction also found to be effective in relevance feedback~\cite{gay2009use}. Type of query document also found to have an influence on similarity model choice~\cite{moreno2015query}, which is also confirmed by our finding. 
Researchers also propose heuristics based~\cite{biggers2014configuring} and search-based~\cite{thomas2013impact,panichella2013effectively} optimization techniques to calibrate IR methods for improved performance. Automatically learning weights while combining different IR models are also proposed in literature~\cite{binkley2014learning,ye2014learning}. 
We complement these works by focusing on similarity model choice for different SE artifacts and demonstrate that an informed choice based on the document features can lead to a better performance.

\section{Threats to Validity}
\label{sec:threats}

From our experimental setup, some threats to the internal validity can arise. Apart from similarity measure, there are some other steps: preprocessing, stopword removal, stemming, etc. that can impact the performance~\cite{panichella2016parameterizing}. We minimize the impact of this threat by applying similar techniques to all the considering models in each step. We also tune each model to its best performing configuration for the corresponding tasks to reduce any parameter configuration bias~\cite{biggers2014configuring,panichella2013effectively}.

As \gh hosts many open-source projects, our category dataset might not be representative enough. To increase the diversity in our project dataset, we use DMOZ Ontology~\cite{dmoz_data_dump}, which is believed to represent the whole Web. However, \gh recently allows users to tag their projects. Though tags are not available for all projects, this tag information could be a possible alternative for DMOZ category.
To curate the dataset for \projecttask, we manually annotated \gh projects, which may impose unwanted bias. To mitigate this, we asked two annotators to annotate separately. We observed an agreement rate of $95\%$ and resolved disagreement by discussion. 



\section{Conclusion}
\label{sec:conc}
In conclusion, we performed an empirical study to understand the interaction between IR-based similarity measures and document types, and observed that model choice has a significant impact on performance for the different types of artifacts as well as mixtures of types. 
With this insight, we further investigate how such an informed choice of similarity measure impacts the performance of IR-based SE tasks. 
In particular, we analyzed two existing tools that leverage diverse SE artifacts. We found that such informed choice of similarity measure indeed leads to improved performance of the SE tools.

%% file: study_subj.tex
\begin{table}[ht]
\centering
\vspace{0.3cm}
\caption{ \textbf{Study Subject}}
\label{tab:subj}
\vspace{0.3cm}
\subfloat[{\small \ProjectTask task}][{ \ProjectTask task}\label{tab:github-project-cumulative-data-stats-category}]
{   
    
    \begin{tabular}{lcccccc}
    \toprule
          &  &  & \textbf{\#Java} & \textbf{\#Method} &  & \textbf{\#Import} \\
          & \textbf{\#Project} & \textbf{\#Category} & \textbf{File} & \textbf{Class} & \textbf{\#API} & \textbf{Package} \\         
    \midrule
    Method-A & 1590  & 78    & 216K  & 4.9M  & 1.5M  & 2.04M \\
    Method-B & 242   & 55    & 14K   & 0.3M  & 0.1M  & 0.12M \\
    \midrule
    Total & 1832  & 112   & 230K & 5.2M & 1.6M & 2.16M \\
    \bottomrule
    \end{tabular}%
}

\subfloat[\small \BugTask task][{ \BugTask task}\label{tab:study-subject-bug}]{
\begin{tabular}{ccccccccc}
    \toprule
         & \textbf{Time Range}  & \textbf{\# bug}  & \multicolumn{2}{p{12.0em}}{\textbf{\# Java files in versions}} & \textbf{\# API}\\
 \textbf{Project} & \textbf{(mm/yy)} & \textbf{reports} & \textbf{median} & \textbf{total} & \textbf{entries} \\
    \midrule
    Birt  & 06/05 -12/13 & 200   & 8770  & 1770K & 957 \\
    Eclipse-UI & 10/01 - 01/14 & 200   & 6141  & 1228K & 1314\\
    JDT   & 10/01 - 01/14 & 500   & 8819  & 4421K & 1329 \\
    SWT   & 02/02 - 01/14 & 200   & 2794  & 559K  & 161\\
    \bottomrule
    \end{tabular}%

}

\end{table}

%% file: table-model-config.tex

\begin{table}[!htbp]
\setlength{\tabcolsep}{2pt}
\renewcommand{\arraystretch}{1.5}
\vspace{0.3cm}
  \centering
  \caption{\textbf{\small Best performing models' configurations}}
    \label{tab:model-best-config}%
  \resizebox{0.95\textwidth}{!}{%
    \begin{tabular}{l|c|cccc|cc|cccc}
    \toprule
     & {\bf VSM} & \multicolumn{4}{c|}{\bf BM25} & \multicolumn{2}{c}{\bf LSI} & \multicolumn{4}{|c}{\bf Word2Vec} \\
    \midrule
               &    Min   &  Min      &       &       &       & Min      & Projected & Min  &            & Window & Vocab \\
     &    DF    &  DF       &  k1   & k2    & b     &  DF      & Dim. & DF   &  Dim. & Size   &  Size \\
    \midrule
    \projecttask & 2     & 2     & 1.5   & 1.5   & 0.75   & 2     & 100   & 5     & 300   & 5     & 18M \\
    \bugtask & 1     & 2     & 1.5   & 1.5   & 0.75   & 15    & 100   & 5     & 100   & 10    & 21.8K \\
    \bottomrule
    \end{tabular}%
    }

  \vspace{0.3cm}
\end{table}%

%% file: table-bug-localization-example-results.tex


  \begin{table}[!htbp]{
  \vspace{0.3cm}
  \centering
  
   \caption{ {Sample results for \bugtask.}}
   \label{tab:bug-result-example}%
   \resizebox{0.95\textwidth}{!}{%
\begin{tabular}{p{31.5em}|p{5.215em}} 
    \toprule
    \textbf{Bug Reports and Fixed File} & \textbf{Rank}\\
    \midrule
    Bug 369884~\cite{bug-report-369884} platform:/plugin/ not used for applicationXMI. ... used for CSS resources or Icons.  I think the ... the applicationXMI parameter. Also the e4 wizard should be adjusted to create the right URI.   \textbf{Fixed File : E4Application.java} ~\cite{E4Application-java-file} & \textbf{BM25=1}  VSM=31  LSI=110 WMD=5983\\
    \bottomrule
    \end{tabular}%
    }
    }
 \end{table}%

